\documentclass[twocolumn,superscriptaddress,PRL]{revtex4-1}
\usepackage{graphicx}
\usepackage{dcolumn}
\usepackage{bm}
\usepackage{color}

\begin{document}


\title{Tunable spin states in two-dimensional magnet CrI$_3$
}

\author{Fawei Zheng}
\email{fwzheng@gmail.com}
\affiliation{Institute of Applied Physics and Computational Mathematics, Beijing, 100088, China}
\affiliation{Beijing Computational Science Research Center, Beijing, 100193, China}

\author{Jize Zhao}
\affiliation{Institute of Applied Physics and Computational Mathematics, Beijing, 100088, China}

\author{Zheng Liu}
\affiliation{Institute for Advanced Study, Tsinghua University, Beijing 100084, China}

\author{Menglei Li}
\affiliation{Department of Physics, Capital Normal University, Beijing 100048, China}

\author{Mei Zhou}
\affiliation{Institute of Applied Physics and Computational Mathematics, Beijing, 100088, China}

\author{Shengbai Zhang}
\email{zhangs9@rpi.edu}
\affiliation{Department of Physics, Applied Physics and Astronomy, Rensselaer Polytechnic Institute, Troy, New York 12180-3590, USA}

\author{Ping Zhang}
\email{zhang\_ping@iapcm.ac.cn}
\affiliation{Institute of Applied Physics and Computational Mathematics, Beijing, 100088, China}
\affiliation{Beijing Computational Science Research Center, Beijing, 100193, China}

\date{\today}

\begin{abstract}
The recent discovery of ferromagnetic single-layer CrI$_3$ creates ample opportunities for studying fundamental properties of atomically-thin magnets. By using first-principles calculations and model analysis, we show that a lateral strain and/or charge doping can have unexpected effects on the magnetic properties of CrI$_3$. In particular, strain tunes the magnetic order and anisotropy: (1) a compressive strain leads to a phase transition from a ferromagnetic insulator to an antiferromagnetic insulator, while (2) a tensile strain can flip the magnetic orientation from off-plane to in-plane. Interestingly, we find that the phase boundary for the first transition is insensitive to charge doping, whereas that of the second one can be significantly modulated by electron doping.
\end{abstract}

\maketitle
It is well known that in a two-dimensional (2D) system, strong fluctuations prohibit the long-range magnetic orders at finite temperature, as dictated by the Mermin-Wagner theorem \cite{MWT}. Interestingly, the single-layer CrI$_3$ (SL-CrI$_3$) was recently found to be a realistic 2D magnet with the Curie temperature Tc $\sim$ 45 K \cite{CrI3Exp}.  The seemingly contradiction could be resolved by considering that the strong spin-orbit coupling (SOC) associated with the heavy anion I$^{-}$ explicitly breaks the SU(2) symmetry of the system and thus Mermin-Wagner theorem does not apply \cite{CrI3SOC}. However, a systematic theoretical study including SOC is still missing. On the other hand, the successful experimental realization of a 2D magnet may provide an exciting new platform for the application of low-dimensional spintronics. This calls for a deeper understanding of its intrinsic physical properties and proposals of practical methods to control its spin states. The most natural methods for atomically thin 2D materials are strain \cite{StrainFeSe1,StrainFeSe2,Strain3} and charge doping \cite{DopingFeSe1,DopingFeSe2,DopingGraphene1,DopingGraphene2}. They may lead to the emergence of new phases and can be used to control their phase transitions. In this letter, we extensively study the strain and charge doping effect. The main results can be illustrated by a phase diagram as shown in Fig. 1(g).  The emergence of new phases and dramatic altering on phase boundaries imply that the strain and charge doping effects are efficient methods to manipulate the spin sates of SL-CrI$_3$.

\begin{figure}[tbp]
\centering
\includegraphics[width=7.0cm]{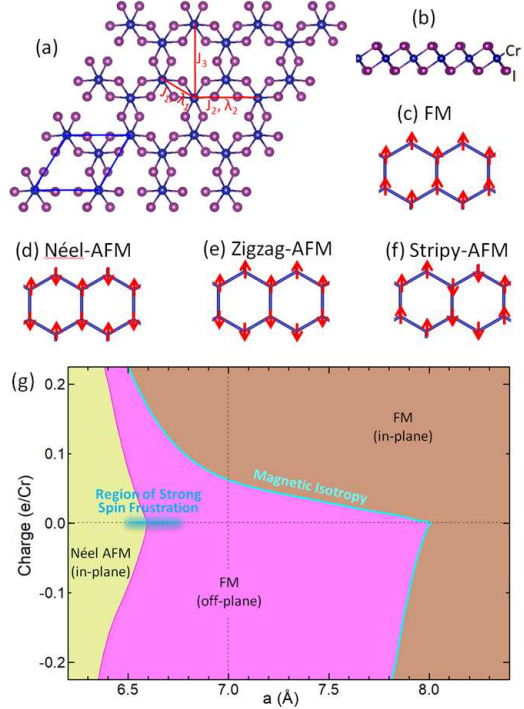}
\caption{\label{fig1} (color online)
    (a) Top and (b) side views of SL-CrI$_3$ atomic structure. Its different magnetic orders are shown in (c-f). Their stabilities are summarized in panel (g).
 The blue parallelogram in (a) shows one unit cell.
	 }
\end{figure}

The density functional theory (DFT) \cite{DFT1,DFT2} calculations in this work explicitly include SOC, based on which the magnetic anisotropy can be quantitatively determined. The calculations are performed by using the projected augmented wave (PAW) method \cite{PAW1,PAW2,PAW3} as implemented in the Vienna ab initio simulation package (VASP) \cite{VASP}. We use the Perdew-Burke-Ernzerhof (PBE) \cite{PBE} type generalized gradient approximation (GGA) in the exchange-correlation functional. An energy cutoff of 227.1 eV is employed. The Brillouin zone is sampled by 8$\times$8$\times$1 k-point mesh primitive cell. For the case of supercell calculations, we keep the k-point density unchanged. The atomic structures are carefully relaxed until the Feynman-Hellman force on each atoms is less than 0.01 eV/\AA. Their stability has been checked by phonon dispersions. To get an accurate density of states (DOS) analysis, we also use the Wannier function techniques as coded in Wannier90 package \cite{WAN}.

To study the SL-CrI$_3$, we adopt the slab model with the vacuum distance 15 \AA. This distance is large enough so the interaction between periodical images can be neglected. The atomic structure of SL-CrI$_3$ is plotted in Fig. 1(a,b). The blue parallelogram in Fig. 1(a) shows the unit cell of a SL-CrI$_3$, containing two Cr atoms and six I atoms. The Cr atoms form a honey-comb lattice as the C atoms in graphene. Six I atoms around each Cr atom form a slightly warped octahedron. It produces a crystal field, which splits the $d$-orbitals into $t_{2g}$ and $e_g$ sub-shells.

The SL-CrI$_3$ has been found experimentally to be ferromagnetism (FM) with an off-plane easy axis \cite{CrI3Exp}.  To validate our calculation, we compare the total energy of the FM order with three other magnetic orders, namely N\'{e}el antiferromagnetism (N\'{e}el-AFM), zigzag anti-ferromagnetism (Zigzag-AFM), and stripy antiferromagnetism (Stripy-AFM) [Fig. 1(c-f)]. For all these magnetic orders, we in addition consider both the in-plane and off-plane spin orientations. The results are summarized in Tab. I. The third column shows the total energy for SL-CrI$_3$ with the lattice parameter $a$ = 7.0 \AA, which is obtained by cell-optimization without external strain. Obviously, the off-plane FM order indeed has the lowest energy. The total energy of off-plane FM order is about 0.7 meV/Cr higher than that of in-plane FM order, which is the magnetic anisotropy energy. The total energies of other magnetic orders, such as N\'{e}el-AFM, Stripy-AFM, and Zigzag-AFM, are at least 11.6 meV/Cr higher than that of off-plane FM. Therefore, they are metastable states.

Considering that these magnetic orders are close in energy, an interesting question is whether the magnetic ground state is tunable by an external perturbation. Our DFT calculations show that the SL-CrI$_3$ is very soft. Its 2D Young’s modulus is only 28 N/m, which is about one order smaller than that of graphene (340 N/m) \cite{GrapheneModulus} and single-layer MoS$_2$ (180 N/m) \cite{MoS2Modulus}. It is even smaller than that of single-layer FeSe (80 N/m from our calculations). This implies that the SL-CrI$_3$ can be easily compressed or stretched, for example, by using the substrate lattice mismatching or piezoelectricity. It is noteworthy that the substrate-induced strain can be as large as 6\% in single-layer FeSe \cite{StrainFeSe1,StrainFeSe2}.  For the two times softer SL-CrI$_3$, we expect that an even larger range of strain modulation can be realized. This leaves plenty room to manipulate the spin state mechanically.

\begin{table}[!hbp]
\begin{tabular}{|c|c|c|c|}
\hline
          & $a$ = 6.3 \AA  & $a$ = 7.0 \AA &  $a$ = 8.2 \AA \\
\hline
in-plane FM & -2.6 & 0.7 & {\bf -0.2} \\
off-plane FM & 0.0 & {\bf 0.0} & 0.0  \\
\hline
in-plane N\'eel-AFM &  {\bf -45.7} & 18.4 & 24.0  \\
off-plane N\'eel-AFM &   -43.2 & 18.5 & 24.9  \\
\hline
in-plane Zigzag-AFM &  -- & 11.6 & --  \\
off-plane Zigzag-AFM &  -- & 11.6 & --  \\
\hline
in-plane Stripy-AFM &  -- & 19.1 & --  \\
off-plane Stripy-AFM &  -- & 19.1 & --  \\
\hline
\end{tabular}
\caption{Total energy ( in meV/Cr ) of SL-CrI$_3$ in different magnetic orders relative to the off-plane FM order. The "--" in the table means that the associated magnetic order is unstable in  DFT calculations. The most stable state for each lattice parameter (a) is shown in bold font.}
\end{table}

\begin{figure}[tbp]
\centering
\includegraphics[width=7.0cm]{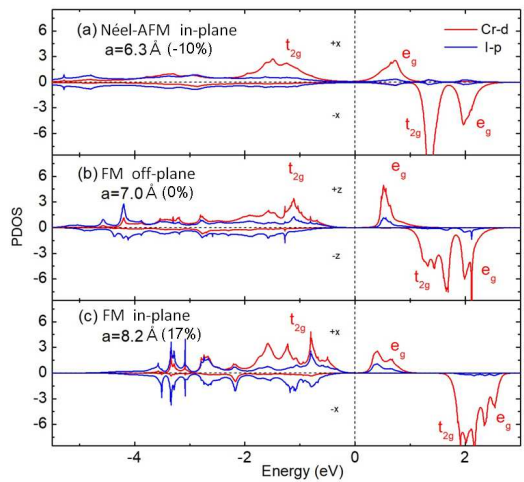}
\caption{\label{fig2}(color online)
	PDOS for in-plane Neel-AFM (a), off-plane FM (b), and in-plane FM(c). The Fermi energy is set to zero, as shown by the black dash line.
}
\end{figure}

We next carry out systematic calculation on the strain effect. In our calculations, the SL-CrI$_3$ is compressed to $a$ = 6.2 \AA\, at most. Our results show that the system has a magnetic phase transition from off-plane FM to in-plane N\'{e}el-AFM at $a$ = 6.6 \AA\,($\sim$ -5.7\%). For example, as shown in the second column of Table I, the total energy of in-plane N\'{e}el-AFM is 45.7 meV/Cr lower than that of off-plane FM, and 2.5 meV/Cr lower than that of off-plane N\'{e}el-AFM at $a$ = 6.3 \AA. Both Zigzag-AFM and Stripy-AFM change to other magnetic orders according to our DFT calculations. Therefore, the in-plane N\'{e}el-AFM is the most stable state at compress strain condition. As for the tensile strain, the system is stretched to $a$ = 8.4 \AA\, at most. There is also a phase transition in this condition. The system would change from off-plane FM to in-plane FM at $a$ = 8.0 \AA\, ($\sim$ 14.3\%). Take $a$ = 8.2 \AA\, for example, as shown in the fourth column of Table I, the total energy of in-plane FM is 0.2 meV lower than that of off-plane FM. The other magnetic orders are either too high in energy or unstable in our DFT calculations.  Therefore, there are only three magnetic orders in the phase diagram. They are in-plane N\'{e}el-AFM, off-plane FM, and in-plane FM.

Furthermore, we calculated the projected density of states (PDOS) of in-plane N\'{e}el-AFM, off-plane FM and in-plane FM phases to get the basic knowledge of their electronic structures. The states near the Fermi energy are mainly composed of Cr-$d$ orbitals, which mix with some I-$p$ orbitals. Thus,  Fig. 2 shows the PDOS on $d$ orbitals for one Cr atom and $p$ orbitals for one I atom. Due to the SOC effect, we can no longer simply use the up and down spins. However, the collinear magnetic moment orientation enables a similar expression, which is the electron spin along with and opposite to the polarization direction. In the cases of in-plane N\'{e}el-AFM and in-plane FM, the magnetic moment can be chosen along any in-plane directions, their energy differences are negligible. We choose the +x direction. Then the DOS is projected to the I-$p$ and Cr-$d$ orbitals with +x and -x spin polarizations.Whereas, in the off-plane FM case, the magnetic moment of Cr atom is along +z direction (the off-plane direction), then the DOS is projected to the I-$p$ and Cr-$d$ orbitals with +z and -z spin polarizations. The PDOS figures of these three phases are similar. There are two groups of Cr-$d$ peaks for +x or +z spin. One group lies below the Fermi energy, the other group lies above the Fermi energy. There are also two groups of Cr-$d$ peaks for -x or -z spin, while both of them stay above the Fermi energy.

We then further project these PDOS to the five $d$ orbitals in octahedron coordinates. The calculation results show that, for the +x and +z spin, the Cr-$d$ peaks below the Fermi energy are composed of $d_{xy}$, $d_{yz}$, and $d_{xz}$ orbitals. They are the three $t_{2g}$ orbitals with three electrons filling in this subshell. This leads to the 3$\mu_B$  net local magnetic moment for each Cr atom. The other two $d$-orbitals belong to $e_g$ subshell, which produces the Cr-$d$ peaks above the Fermi energy.  For the -x and -z spin, both the $t_{2g}$ and $e_g$ subshells stay above the Fermi energy, they are responsible for the two groups of Cr-$d$ peaks above the Fermi energy. The PDOS figures show that all the three phases are insulators. The insulating gap lies between the occupied $t_{2g}$ levels and the empty $e_g$ levels and thus is primarily determined by the crystal-field splitting. Our energy band calculations show that the band gap in Fig. 2(a) is 0.38 eV, which is smaller than the 0.88 eV gap value in Fig. 2(b). When the lattice is compressed, the distances between atoms are smaller, thus the crystal field effect is stronger. It agrees with the PDOS peak position moving trend. In Fig. 2(a), the energy difference between PDOS peaks bellow and above the Fermi energy is larger than that in Fig. 2(b). This increases the band gap in principle. While as atoms are close to each other, the overlap between their orbitals increases. This leads to a larger band width, and effectively reduces the band gap. All together, the band gap is reduced. When the lattice is expanded, we can see that the band gap in Fig. 2 (c) is 0.65 eV, which is also smaller than that in Fig. 2(b). In this case, the larger distance between atoms weakens the crystal field and reduces the orbital overlaps. The whole effect reduces the band gap.

\begin{figure}[tbp]
\centering
\includegraphics[width=8.6cm]{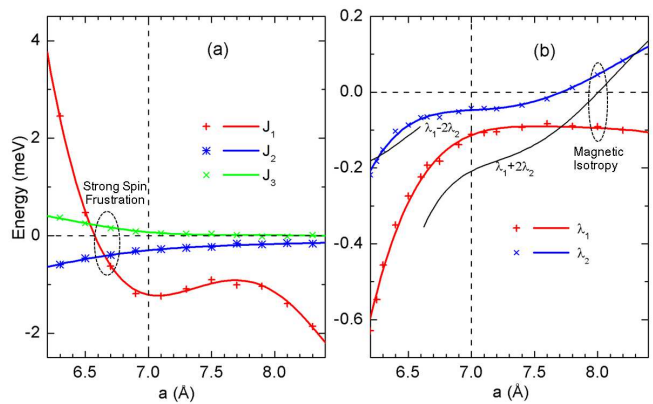}
\caption{\label{fig3}(color online)
	The fitted (a) Heisenberg exchange parameters and (b) anisotropy
exchange parameters as functions of lattice parameter.
}
\end{figure}

To study the mechanism of the phase transitions induced by strain in detail, a model Hamiltonian could be very helpful. Actually, Wang {\it et al.} have proposed one to explain the magnetism of SL-CrI$_3$ at ambient pressure \cite{CrI3DFT}. They considered the first, second and third nearest neighbor Heisenberg exchange integrations and found that the first and second nearest neighbor Heisenberg integrations are negative, while the third one is positive.  Moreover, the latter one is much smaller than the former two. This model successfully explains the ferromagnetism of SL-CrI$_3$ without external strain. However, it cannot explain the intrinsic magnetic anisotropy. Then, Lado and Fernandez-Rossier proposed a nearest neighbor XXZ model \cite{CrI3SOC}. They found that the anisotropy arises predominantly from the SOC in I atoms via the anisotropic symmetric superexchange. The XXZ model is capable of explaining the magnetic anisotropy at zero strain.  To include the strain effect on the magnetism of SL-CrI$_3$, in the present work, we extend the three-neighbor Heisenberg model and the nearest neighbor XXZ model to a model Hamiltonian as,
\begin{eqnarray}
H=&&\sum_{nn}(J_1\vec{S_1}\cdot\vec{S_2}+\lambda_1S_1^zS_2^z)+\sum_{2nn}(J_2\vec{S_1}\cdot\vec{S_2}+\lambda_2S_1^zS_2^z)\nonumber\\
&&+\sum_{3nn}J_3\vec{S_1}\cdot\vec{S_2}\, .
\end{eqnarray}
In this model, we consider the first, second, and third nearest neighbor Heisenberg exchange integrations, as well as the first and second anisotropy exchange integrations. The associated Heisenberg exchange integration parameters are $J_1$, $J_2$, and $J_3$, and the anisotropy exchange integrations are $\lambda_1$ and $\lambda_2$, respectively.  If we only keep $\lambda_1$ in our model Hamiltonian, the model could describe the magnetic behaviors near zero strain accurately, while it fails to explain the magnetic properties of CrI$_3$ under a large tensile  strain. Moreover, the fitted parameter $\lambda_1$ would have very large uncertainty in the absence of $\lambda_2$. Therefore, the parameter $\lambda_2$ is indispensable. We then further check a model with $\lambda_1$, $\lambda_2$, and the third nearest neighbor anisotropy exchange integration $\lambda_3$. The model nicely reproduces the magnetic properties of CrI$_3$ in the whole range of lattice parameters. However, the fitted $\lambda_3$ is at least one order smaller than $\lambda_2$, hence we can safely neglect it. In a word, the accurate and minimal model contains $J_1$, $J_2$, $J_3$, $\lambda_1$, and $\lambda_2$.

These model parameters are fitted from our DFT total energy calculations. We firstly choose a group of different spin configurations, and then calculate their total energies.  After that, linear regressions are carried out to obtain the values of model parameters.  The fitted parameters are shown in Fig. 3. The left panel shows three Heisenberg exchange integrations. We can see that the first nearest neighbor parameter $J_1$ has large amplitude and changes its sign at $a$ = 6.5 \AA. The parameter $J_1$ contains two kinds of contributions, they are the direct and indirect Heisenberg exchange integrations, respectively. The Cr-Cr direct exchange integration is positive, which favors AFM. Three unpaired electrons fill the up-spin (+x or +z) $t_{2g}$ orbitals, composing a closed shell. Due to the different symmetries, electrons from the nearest neighbor Cr atoms can not hop to the up-spin $e_g$ orbitals.  They can only hop to the down-spin $t_{2g}$  orbitals, which lead to the positive sign of the direct exchange integrations eventually. On the other hand, due to the near $\pi$/2 bond angle of Cr-I-Cr, the indirect exchange integration is negative, which favors FM. Then, the competition between the direct and indirect integrations evokes the complex behaviors of $J_1$. The direct integration depends on the distance between two nearest Cr atoms, which increases quickly with the decreasing of lattice parameter. While, the indirect integration depends on both the Cr-I bond length and their relative orientations. When the SL-CrI$_3$ is compressed, the Cr-I-Cr angle slightly decreases, while the bond length is nearly unchanged. Therefore, the indirect integration is less sensitive to the lattice compression. If the system is heavily compressed, the direct integration becomes the dominant contribution, as a result, the total integration $J_1$ is positive. When the system is not heavily compressed, the indirect integration is the dominant contribution, then the $J_1$ is negative. The second and third nearest neighbor Heisenberg integration parameters $J_2$ and $J_3$ are also shown in Fig. 3(a). We can see that their amplitudes are relatively small, and decrease with increasing lattice parameter. The distances between a Cr atom to its second and third nearest neighboring Cr atoms are very large ( $>$ 6 \AA), thus the direct integrations are negligible, only indirect integrations contribute to $J_2$ and $J_3$. The external strain only smoothly alters their amplitudes while keeps their signs unchanged.

The system contains spin frustration, which can be shown by analyzing the lattice structure and model parameters. The honeycomb lattice has two sub-lattices, noted as A and B. The Heisenberg term $J_2$ acts within each sub-lattice. The negative sign of $J_2$ indicates that it is energetically favorable for atoms to align ferromagnetically within each sublattice. On the other hand, the $J_3$ term is always positive, preferring an antiferromagnetic configuration between the two sublattices. The $J_1$ term, which can be positive or negative, determines the magnetic properties of CrI$_3$ crucially. For a negative $J_1$, there are no spin configurations to minimize the energy on all the bonds, which is known as frustration. It is highly expected that strong frustrations can lead to exotic states of matter, such as spin liquids. Therefore, CrI$_3$ offers a promising chance to study those nontrivial phenomena experimentally.

The fitted spin anisotropy parameter $\lambda_1$ and $\lambda_2$ are shown in Fig. 3(b). The parameter $\lambda_1$ has relatively high amplitude, and it is always negative. The parameter $\lambda_2$ is negative for $a <$ 7.2 \AA, and changes to positive for $a >$ 7.2 \AA. As we can see from Fig. 1, each Cr atom has three nearest neighbor Cr atoms and six second nearest neighbor Cr atoms. Therefore, the total anisotropy is proportional to a combination of $\lambda_1$ and $2\lambda_2$. For the case of N\'{e}el-AFM, the total anisotropy is proportional to $\lambda_1-2\lambda_2$. The system would become off-plane for positive values of $\lambda_1-2\lambda_2$, and in-plane for negative values of $\lambda_1-2\lambda_2$.  From Fig. 3(b), we see that the value of $\lambda_1-2\lambda_2$ is negative for the range of lattice parameter in N\'{e}el-AFM phase. Therefore the magnetic orientation is in-plane. For the case of FM, the total anisotropy is proportional to $\lambda_1+2\lambda_2$. The system is off-plane and in-plane for negative and positive values of $\lambda_1+2\lambda_2$, respectively. From Fig. 3(b), we know that the $\lambda_1+2\lambda_2$ is positive for  $a <$ 8.0 \AA, while, it is negative for $a >$ 8.0 \AA. Therefore, the system is off-plane FM for $a <$ 8.0 \AA, and it is off-plane FM for $a >$ 8.0 \AA. The sign change is mainly introduced by the variation of $\lambda_2$, thus the phase transition at $a$ = 8.0 \AA\, is driven by the anisotropy of the second nearest neighbor interaction.

To confirm the validation of our model Hamiltonian, we also perform a cluster mean-filed analysis. One hexagon ring is decoupled from the honeycomb lattice, and the surroundings are taken as effective fields. The Hamiltonian of such a cluster can be numerically diagonalized easily,  and the expectation values of the spins on the cluster are readily available. These expectation values are related to the effective fields by the symmetry transformation associated with the lattice and therefore they can be computed self-consistently. By using this method, the phases and critical points agree well with those obtained by DFT calculations as shown in Fig. 1(c). This provides further evidence to support our DFT calculations.

Besides strain, another common effect in epitaxial single layer is charge doping. Usually, the substrate and epitaxial single layer are composed of different materials. They have different Fermi energies. Then the electrons or holes would transfer from substrate to the epitaxial layer, and vice versa. Actually, the interface induced charge transfer has been widely used to tune the properties of epitaxial layers, such as the single-layer FeSe \cite{DopingFeSe1} and graphene \cite{DopingGraphene1,DopingGraphene2}. Besides substrate, the charge doping to an ultra-thin layer material can also be introduced effectively by using electrical methods or by deposition of atoms. For example, the potassium atoms are used extensively to doping electrons to FeSe layer \cite{DopingFeSe2,DopingFeSe3,DopingFeSe4,DopingFeSe5}. Therefore, additional charge can be easily doped to a SL-CrI$_3$ to tune its properties. Here, we also study this effect.

The calculation results of charge doping effect are demonstrated in the phase diagram Fig. 1(g).  We can see that the AFM-FM phase transition is robust as the critical strains are only slightly changed by charge doping. The off-plane FM phase extendes slightly to the left side for both the electron and hole dopings. Therefore, both charge dopings stabilize the FM phase. This agrees with the previous study on the energy difference between FM and AFM SL-CrI$_3$ without strain \cite{CrI3DFT}. The phase transition between in-plane and off-plane FM phases in Fig. 1(g) behaves rather differently. Specially, the hole doping only slightly extends the in-plane FM phase to its left side, while the electron doping extends the in-plane FM phase to its left side dramatically. Under electron doping, the phase boundary crosses the vertical line at 7 \AA. The crossing point locates at 0.06 e/Cr, which is not very large comparing to the doping density in single-layer FeSe experiments \cite{DopingFeSe2,DopingFeSe3,DopingFeSe4}. Therefore, the charge doping can also be used to tune the magnetic anisotropy and spin orientation in SL-CrI$_3$.

In conclusion, we have studied the effects of the strain and charge doping on SL-CrI$_3$. Our calculations show that the SL-CrI$_3$ has a rich magnetic phase diagram, including in-plane AFM, off-plane FM, and in-plane FM phases. The SL-CrI$_3$ is found to be very soft. Therefore strain can be easily applied to it.  Under compressive strain, the system changes from off-plane FM phase to in-plane N\'{e}el-AFM phase. Their phase boundary is found to be very robust to both electron and hole dopings. When the system is under tensile strain, the SL-CrI$_3$ would change to in-plane FM phase.  The hole doping only slightly changes the phase boundary between in-plane and off-plane FM phases, while the electron doping changes the phase boundary dramatically. We have proposed a model Hamiltonian to explain the phase transitions. Our analysis shows that the first nearest neighbor Heisenberg exchange integration changes its values dramatically, and triggers the FM-AFM phase transition. Whereas, the second nearest neighbor anisotropy exchange integration changes its sign and values with tensile strain and rules the phase transition between in-plane and off-plane FM phases. Considering the continuously tunable nature of strain and charge doping in experiment, our present findings on extraordinarily strong couplings of spin with strain and electron doping enable the SL-CrI$_3$ to be a convenient platform for the study of spin frustration and Mermin-Wagner theorem, as well as a promising building block for spintronics.

This work was supported by the National Natural Science Foundation of China (Grants No. 11474030 and No. 11625415), National Basic Research Program of China (973 program, Grants No. 2015CB921103), and President Fund of China Academy of Engineering Physics (Grants No. YZJJLX2016010).


\begin{thebibliography}{99}
\bibitem{MWT} N. D. Mermin, and H. Wagner, Phys. Rev. Lett. {\bf 17}, 1133 (1966).
\bibitem{CrI3Exp} B. Huang, G. Clark, E. Navarro-Moratalla, D. R. Klein, R. Cheng, K. L. Seyler, D. Zhong, {\it et al.}, Nat., {\bf 546}, 270 (2017).
\bibitem{CrI3SOC} J. L. Lado, and J. Fernandez-Rossier, 2D Mater. {\bf 4}, 3 (2017).

\bibitem{StrainFeSe1} P. Zhang, X.-L. Peng, T. Qian, P. Richard, X. Shi, J.-Z. Ma, B. B. Fu,  {\it et al.},  Phys. Rev. B, {\bf 94 }, 104510 (2016).
\bibitem{StrainFeSe2} R. Peng, H. C. Xu, S. Y. Tan, H. Y. Cao, M. Xia, X. P. Shen, Z. C. Huang, {\it et al.}, Nat. Commun. {\bf 5}, 5044 (2014).
\bibitem{Strain3} M. Zhou, W. Duan, Y. Chen, and A. Du, Nanoscale, {\bf 7}, 15168 (2015).

\bibitem{DopingFeSe1} Q.-Y. Wang, Z. Li, W.-H. Zhang, Z.-C. Zhang, J.-S. Zhang, W. Li, H. Ding,  {\it et al.}, Chin. Phys. Lett. {\bf 29}, 037402 (2012).
\bibitem{DopingFeSe2} Y. Miyata, K. Nakayama, K. Sugawara, T. Sato, T. Takahashi, Nat. Mat. {\bf 14}, 775 (2015).

\bibitem{DopingGraphene1} J. E. Lee, G. Ahn, J. Shim, Y. S. Lee, and S. Ryu, Nat. Commun. {\bf 3}, 1024 (2012).
\bibitem{DopingGraphene2} J. Shim, C. H. Lui, T. Y. Ko, Y.-J. Yu, P. Kim, T. F. Heinz, and S. Ryu, Nano Lett, {\bf 12}, 648 (2012).

\bibitem{DFT1} P. Hohenberg, and W. Kohn, Phys. Rev. {\bf 136}, B864 (1964).
\bibitem{DFT2} W. Kohn, and L. J. Sham, Phys. Rev. {\bf 140}, A1133 (1965).
\bibitem{PAW1} P. E. Blöchl, Phys. Rev. B, {\bf 50}, 17953 (1994).
\bibitem{PAW2} G. Kresse, and D. Joubert, Phys. Rev. B, {\bf 59}, 1758 (1999).
\bibitem{PAW3} P. E. Blöchl, C. J. Först, and J. Schimpl,  B. Mater. Sci. {\bf  26}, 33 (2003).
\bibitem{VASP} G. Kresse, and J. Furthmüller,  Phys. Rev. B, {\bf 54}, 11169 (1996).
\bibitem{PBE} J. P. Perdew, K. Burke, and M. Ernzerhof, Phys. Rev. Lett. {\bf 77}, 3865 (1996).
\bibitem{WAN} A. A. Mostofi, J. R. Yates, G. Pizzi, Y. S. Lee, I. Souza, D. Vanderbilt, and N. Marzari, Comput. Phys. Commun. {\bf 185}, 2309 (2014).



\bibitem{GrapheneModulus}  A. Politano, and  G. Chiarello, Nano Res., {\bf 8}, 1847 (2015).
\bibitem{MoS2Modulus} S. Bertolazzi, J. Brivio, and A. Kis, ACS Nano, {\bf 5}, 9703 (2011).
\bibitem{CrI3DFT} H. Wang, F. Fan, S. Zhu, and H. Wu, EPL, {\bf 114}, 47001 (2016).

\bibitem{DopingFeSe3} C.-L. Song, H.-M. Zhang, Y. Zhong, X.-P. Hu, S.-H. Ji, L. Wang, K. He, X.-C. Ma, and Q.-K. Xue, Phys. Rev. Lett. {\bf 116}, 157001 (2016).
\bibitem{DopingFeSe4} M. Ren, Y. Yan, X. Niu, R. Tao, D. Hu, R. Peng, B. Xie, J. Zhao, T. Zhang, and D.-L. Feng, Sci. Adv., {\bf 3}, 1603238 (2017).
\bibitem{DopingFeSe5} F. Zheng, L.-L. Wang, Q.-K. Xue, and P. Zhang, Phys. Rev. B {\bf 93}, 075428 (2016).

\end{thebibliography}
\end{document}